\documentstyle[prl,aps,multicol,epsf]{revtex} 
\tighten 
\input epsf 
 
\newcommand{\D}{{\rm d}} 
\newcommand{\I}{{\rm i}} 
\newcommand{\Tr}{{\rm Tr}} 
\renewcommand{\Re}{{\rm Re}} 
\renewcommand{\Im}{{\rm Im}}

\firstfigfalse

\begin{document} 
\title{Magnetoconductance of a two-dimensional metal in the 
presence of spin-orbit coupling} 
\author{P. Schwab$^{(1)}$ and R. Raimondi$^{(2)}$} 
\address{ 
 $^{(1)}$Institut f\"ur Physik, Universit\"at Augsburg, D-86135 Augsburg \\ 
 $^{(2)}$NEST-INFM e Dipartimento di Fisica, 
 Universit\`a di Roma Tre, Via della Vasca Navale 84, 00146 Roma, Italy } 
\date{\today} 
\maketitle 
\begin{abstract} 
We show that in the metallic phase of a two dimensional electron gas the 
spin-orbit coupling due to structure inversion
asymmetry leads to a characteristic anisotropy in the magnetoconductance.
Within the assumption that the metallic phase can be described by a Fermi liquid,
we compute the conductivity in the presence of an in-plane magnetic field.
Both the spin-orbit coupling and the Zeeman coupling with the magnetic field
give rise to two spin subbands, in terms of which most of the transport
properties can be discussed. 
The strongest conductivity anisotropy occurs for
Zeeman energies of the order of the Fermi energy corresponding to
the depopulation of the upper spin subband. The energy scale associated
with the spin-orbit coupling controls the strength of the effect.
More in particular, we find that the detailed behavior and 
the sign of the anisotropy depends on the underlying 
scattering mechanism.
Assuming small angle scattering to be the dominant scattering mechanism our
results agree with recent measurement on Si-MOSFET's  
in the vicinity of the metal-insulator transition.
\end{abstract} 
\begin{multicols}{2}
\section{Introduction}
In recent years
the transport properties of two-dimensional (2D) electron systems in
Si-MOSFETs and semiconductor heterostructures have been the subject of
a great deal of theoretical and experimental activity.
One reason of this is the possibility of a metal-insulator transition
(MIT) as the density of the 2D system is varied\cite{kravchenko95}.
A ``critical'' density identified by  a weakly temperature dependent resistivity 
separates an ``insulating'' region from a ``metallic'' one.
In the latter, upon lowering the temperature roughly below $1$K, 
the resistivity drops by almost one order of
magnitude, whereas in the former the resistivity increases quickly.
The origin of the MIT is still unclear and there is a wide debate in the
literature concerning the relevance of the possible mechanisms. 
For recent reviews on the subject one may see Ref.\cite{abrahams00} and \cite{altshuler00}.

On the insulating side the resistance resembles Efros-Shklovskii hopping\cite{mason95}, 
which points towards the relevance of localized states and a soft gap 
in the density of states; also the thermopower has been interpreted in terms of an
Anderson insulator\cite{fletcher00}.
From this point of view Anderson localization seems 
to be relevant for the transition. 
Whereas
the standard single parameter scaling theory predicts no
Anderson transition
in two dimensions, 
on the basis of the existing 
scaling theory of interacting electrons  
a transition cannot be excluded
\cite{finkelstein83,castellani84,dobro97,castellani98,punnoose01}.
Besides, 
several properties of the metallic side, including the low density, i.e., 
the large value of
the electron gas parameter 
$r_s$, the strong increase of the electron $g$-factor\cite{okamoto99,kravchenko00}, 
the anomalous magnetoresistance \cite{simonian97,pudalov97},
and a vanishing compressibility at the critical 
density\cite{dultz00} can indeed be seen as hints for the relevance 
of the electron correlations; quantum corrections to the conductivity on the other hand seem to 
be only small\cite{pudalov99,simmons99,senz00}, at least in the range of temperatures
where the main drop of the resistivity occurs\cite{altshulermartin00}.

The main and still open questions are:
(1) Is there a real quantum phase transition, or is just a temperature dependent
crossover phenomenon observed?
(2) What is the nature of the insulating state? 
(3) Can the metallic phase be described in terms of the standard Fermi-liquid theory?

In this paper we will not touch point one and two. 
Concerning point three, there is accumulating 
evidence that the answer is ``yes'': 
at the lowest temperatures and weak magnetic fields 
standard weak localization physics
has been observed in a number of different systems. 
 
In order to obtain a better understanding of the materials, different aspects 
have to be investigated.
We study the influence of spin-orbit coupling on the magnetoconductance.

We assume that the metallic phase can be described in terms of the standard Fermi-liquid
 theory,
and therefore we calculate transport properties in the framework of the
 Drude-Boltzmann theory.
In Si-MOSFET's with low electron density the major spin-orbit term 
is believed to be due to the
 lack
of inversion symmetry of the confining potential. Originally, there has been the
 suggestion that 
the spin-orbit coupling may be responsible for the observed metallic behavior of the
 resistivity\cite{pudalov97a}.
Although at present such a view is not confirmed by the experiments,
it is certainly useful to assess the relevance of the spin-orbit coupling in these systems as well as
to understand whether it is related to 
the puzzling magnetoresistance in a parallel field. In this respect,
Chen et al\cite{chen00} pointed out that such a spin-orbit coupling
would induce an anisotropy in the magnetoresistance.
The resistivities measured  parallel and perpendicular to
the in-plane magnetic  field are different.
Such an anisotropy has been observed in Si-MOSFETs\cite{pudalov00}.

However, while the original theoretical proposal in Ref.\cite{chen00} 
has been formulated in the  localized limit, 
the experimental results have been obtained 
near and across the MIT.
For this reason a direct comparison between theory 
and experiment is difficult.
In a recent paper\cite{leadbeater01} we have provided a theory valid in the 
metallic regime. 
We were able to reproduce correctly a number of the experimental findings.
The only drawback of our theory was the wrong sign of the effect, 
and we speculated that this could be due to the oversimplified assumption of 
a pure $s$-wave impurity scattering.
It is also worth mentioning that anisotropic magnetoresistance has been
observed in 
$GaAs$ electron systems\cite{khrapai00}, and $GaAs$ hole systems\cite{papadakis99,noh01}.
In these systems, the interpretation is complicated by the intrinsic anisotropy
associated with the crystal structure.

In this paper, we study how the form of the microscopic scattering potential affects
the anisotropy of the magnetoconductance in the presence of the spin-orbit coupling.
In particular, we extend our previous analysis to include the case of  small angle
 scattering.
The layout of our paper is the following.
In section II we introduce the model and make a first analysis by means of 
 the Boltzmann equation. We adopt   the relaxation-time approximation
assuming a relaxation rate independent from the position on 
the Fermi surface. 
In section III,  starting  from the Kubo formula we  calculate the 
conductivity within a Green function approach. As a first step, we avoid
to  specify a microscopic scattering mechanism
and we simply introduce, by hand,
a lifetime in the one-particle Green function. In so doing we neglect vertex corrections
but take into account
level broadening and inter-band transitions. 
As it will be clear in the following, while vertex corrections may give rise to
significant changes in the conductivity, the  inclusion of level broadening
affects only weakly  the results.
A more sophisticated analysis is carried out in
sections IV and V, where  we will specify a microscopic mechanism responsible for the
finite conductivity, namely elastic scattering. 
In section IV we will assume s-wave scattering. The Green functions 
and vertex corrections  will be
calculated in the self-consistent Born approximation.
In section V we will explore the consequences of a strongly
angle dependent scattering potential.
In the limit of strong forward scattering
we will find that the 
sign of the anisotropy is consistent with the experiments.  Finally,
section VI will contain our conclusions.
\section{The model and the relaxation time approximation} \label{secRelax}
We start from the model Hamiltonian 
\begin{equation}
\label{1}
H= \frac{p^2}{2m}+\alpha {\mathbf {\sigma}}\cdot {\bf p}\wedge{\bf e}_z -
\frac{1}{2}g\mu_B{\mathbf \sigma}\cdot {\bf B}
\end{equation}
where  $\alpha$ is a parameter describing the spin-orbit coupling due to the 
confinement field\cite{rashba84}, 
${\bf {\sigma}}$ is the  Pauli matrices  vector, and
${\bf e}_z$ is a unit vector perpendicular to the 2D system.
As pointed out recently by Winkler\cite{winkler00}, 
this so-called Rashba model applies for 2D electron systems 
(like Si-MOS in Ref.\cite{pudalov00}), 
but not for heavy-hole states which are relevant e.g. in Si/SiGe quantum wells.
To simplify the notation we introduce the Zeeman energy $\omega_s
=\frac{1}{2}g\mu_B B$ (note that this definition differs by a factor of $2$ from the
standard one).
The $g$-factor in nSi-MOS structures is near two but is growing with decreasing electron 
density\cite{okamoto99}.
 In the presence  of spin-orbit coupling and magnetic field,
the original  spin degenerate band  splits into two bands with energy dispersion
\begin{eqnarray}
\label{2}
E_{\pm}({\bf p})&  =&  p^2/2m \pm \Omega({\bf p} )\\ 
\Omega({\bf p} )&  = & \sqrt{(\alpha p_y -\omega_s)^2 + (\alpha p_x)^2}
.\end{eqnarray}
We have chosen the direction of the in-plane magnetic field as the
{\bf x}-axis.
According to the  Boltzmann transport theory within the relaxation time approximation,
the conductivity tensor at zero temperature is
\begin{equation}
\label{4}
\sigma_{ij } =
e^2 \sum_{{\bf  p},s } v_{ s}^i v_{s }^j
 \tau_{{\bf p}, s}
 \delta\left( E_s({\bf p })  -\mu \right)
 ,\end{equation}
where $s=\pm$ is the band index and the velocities are 
${\bf v}_\pm = \nabla_{\bf p} E_\pm({\bf p}) $.
By further assuming  a constant relaxation rate 
$\tau_{{\bf p}, s} = \tau_{\rm tr}$,
the conductivity is proportional to the product
of the density of states, $N_s (\mu )$,  times a Fermi surface average of the Fermi velocities,
\begin{equation}
\label{5}
\sigma_{ij} = e^2 \tau_{\rm tr}
\sum_s N_{s}(\mu )
\langle v^i_s v^j_s  \rangle_{FS_s} .
\end{equation}
We note also that, 
after integrating by parts Eq.(\ref{4}), the conductivity  may be 
expressed in terms of the effective mass tensor,
\begin{equation} \label{eqSigmaMass}
\sigma_{ij} = e^2 \tau_{\rm tr}
\sum_{{\bf p}, s}
{ \partial^2 E_s({\bf p}) \over \partial p_i \partial p_j }
\Theta(\mu - E_s({\bf p} )).
\end{equation}
In the absence of spin-orbit scattering this relation becomes simply
$\sigma = 2 e^2 \tau_{\rm tr} n/m$, with $n$  the electron density.
We assume that the density does not change with the magnetic field and therefore we have to 
adjust the chemical potential. As long as both bands are occupied, the chemical
potential is nearly constant, $\mu(\omega_s ) \approx \epsilon_F$. For high fields 
when the upper band
is depopulated it decreases as $\mu (\omega_s )  \approx 2\epsilon_F - \omega_s$.
The chemical potential in the absence of magnetic field is here denoted by $\epsilon_F$.

From Eqs.(\ref{5}) and (\ref{eqSigmaMass}) it is apparent that the transport time drops
when we consider the conductivity relative to its value in the absence of 
magnetic field
and spin orbit scattering, $\sigma_0 =  e^2 N_0 v_F^2 \tau_{\rm tr} $.
For the same reason,  the relative anisotropy in the conductivity,
$\Delta \sigma = {(\sigma_{xx} - \sigma_{yy} )/ \sigma_0 }$,
does not depend on the scattering time $\tau_{\rm tr} $ and 
may be determined from the simple knowledge of the band structure.
This observation makes clear the main goal of our analysis, i.e. to what extent
 we can understand the
small anisotropy in the conductance without
making any assumption about the origin of the strong magnetic field dependence of
the conductivity itself.
\begin{figure}
\noindent
\begin{minipage}[t]{0.98\linewidth}
{\centerline{\epsfxsize=7cm\epsfbox{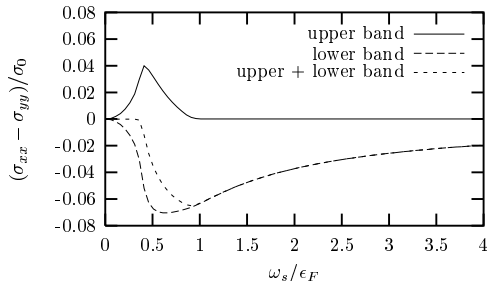}}}
\caption{$\Delta\sigma$ versus magnetic field 
($\omega_s= g\mu_B B/2$) at fixed spin-orbit 
coupling
$\alpha p_F =0.4 \epsilon_F$ as obtained within the relaxation time approximation
with a single scattering rate $\tau_{\rm tr}$.}
\label{figr1}
\end{minipage}
\end{figure}
The evaluation of the conductivity via Eq.(\ref{5}) or (\ref{eqSigmaMass}) is analytically
nasty due to the presence of the square root in the energy band dispersion
 (cf. Eq.(\ref{2})).
We begin then by presenting the result of the numerical integration of Eq.(\ref{eqSigmaMass}) in
Figure \ref{figr1} which shows $\Delta \sigma $ for a special value of the spin-orbit term 
as a function of the magnetic field.
We have also plotted the contributions for the two bands separately, and it is 
seen that at weak magnetic field there is a strong cancellation from the 
contributions of both bands.
At large field the upper band depopulates and therefore does not contribute 
to the conductivity or its anisotropy.

Some more insight is obtained by considering Eq.(\ref{eqSigmaMass}) in the weak and strong field
limits where analytical progress is possible. 
First we note that  $\Delta \sigma $ must be zero in the limits of very weak and 
very strong magnetic field, when the Fermi surface becomes rotational symmetric.
When $\omega_s > \alpha p_F$ the dispersion of the two bands
is approximately given by
\begin{equation}
\label{7}
E_{\pm}({\bf p } ) = p^2/2m \pm \left( 
 \omega_s - \alpha p_y + (\alpha p_x)^2/2 \omega_s
\right)
,\end{equation}
so that the effective mass in the $x$-direction, i.e. parallel to the magnetic field,
is modified according to
\begin{equation}
\label{8}
{1\over m_{x, \pm} } = {1\over m} \left(
1 \pm {m \alpha^2 \over \omega_s } \right)
.\end{equation}
For $\omega_s > \epsilon_F$, when only one band is occupied,
the anisotropy in the conductivity is thus
proportional to $1/m_x - 1/m$ and explicitly given by
\begin{equation}
\label{9}
{\sigma_{xx}-\sigma_{yy}\over \sigma_0} =
 - {1\over 2} {(\alpha p_F)^2 \over \omega_s \epsilon_F}
.\end{equation}
For $\omega_s < \epsilon_F$,
when both bands contribute, there is a partial cancellation:
\begin{eqnarray}
\label{10}
{\sigma_{xx}- \sigma_{yy} \over \sigma_0 } 
&=& -{1\over 2}{(\alpha p_F)^2 \over \omega_s \epsilon_F}{ n_- - n_+ \over n}\\
&\approx& -{1\over 2}{(\alpha p_F)^2 \over \epsilon_F^2}
,\end{eqnarray}
since the electron densities in the lower and upper band 
are $n_{\pm} \approx n ( 1 \mp \omega_s/\epsilon_F)/2$.

\begin{figure}
\noindent
\begin{minipage}[t]{0.98\linewidth}
{\centerline{\epsfxsize=7cm\epsfbox{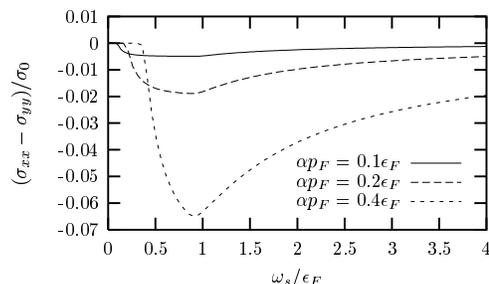}}}
\caption{$\Delta\sigma$ versus magnetic field 
as obtained within the relaxation time approximation
for various strengths of the spin-orbit coupling. }
\label{fig2}
\end{minipage}
\end{figure}

In order to give the results for  the low field limit,
we first write (\ref{eqSigmaMass}) explicitly for the given band dispersion
\begin{eqnarray}
\label{12}
\sigma_{xx} - \sigma_{yy} & = &
e^2 \tau_{\rm tr} \alpha^2
\sum_{\bf  p }{ (\alpha p_x )^2 -(\omega_s - \alpha p_y)^2 \over 
\Omega^3 } \nonumber \\
&& \times \left[ \Theta(\mu- E_-({\bf p}) ) - \Theta( \mu - E_+({\bf p} ) )
\right]
.\end{eqnarray}
When $\Omega \ll \epsilon_F$ one may then approximate the difference of
the step functions by a delta function which leads to
\begin{eqnarray} \label{eq13}
{\sigma_{xx} - \sigma_{yy} \over \sigma_0}& =  &{1\over 2}{(\alpha p_F)^2 \over \epsilon_F^2 }
\\
&& \times \int_0^{2\pi} {\D \theta \over 2 \pi}
{ 
[\alpha p_F \cos(\theta)]^2- [\omega_s - \alpha p_F \sin(\theta)]^2\over \Omega^2}
.\nonumber \end{eqnarray}
Evaluating the integral one arrives at
\begin{equation} \label{eq14}
{\sigma_{xx} - \sigma_{yy} \over \sigma_0}= 
{1\over 2} {(\alpha p_F)^2 \over \epsilon_F^2 }
{ 1- (\omega_s/\alpha p_F)^2 \over (\omega_s /\alpha p_F)^2 }
\Theta( \omega_s/\alpha p_F - 1 )
.\end{equation}
It is interesting to note that
(\ref{eq13}) is identical to the Fermi surface average of the anomalous part of the
velocity ${\bf v}_{\pm} = {\bf p}/m \pm \alpha {\bf e}_{\bf p}$,
\begin{equation} \label{eq15}
\sigma_{xx} - \sigma_{yy} = 2 e^2 N_0 \tau_{\rm tr} \alpha^2 \langle 
( {\bf e}_{\bf p } \cdot {\bf e}_x)^2-
( {\bf e}_{\bf p } \cdot {\bf e}_y)^2
\rangle_{FS}
\end{equation}
with ${\bf e}_{\bf p} = (\alpha {\bf p} - \omega_s {\bf e}_y )/\Omega$.
At low magnetic field the vector ${\bf e}_{\bf p}$ makes a full rotation, when averaging
 over
the Fermi surface so that the two dot products averaged over the Fermi surface,
 in Eq.(\ref{eq15}),
cancel each other. The anisotropy of the conductivity vanishes.
At large field ${\bf e}_{\bf p}$ becomes locked in $y$-direction. The conductivity
 becomes  then
anisotropic at a magnetic field value $\omega_s =\alpha p_F$.
Fig.\ref{fig2} shows 
the numerically determined anisotropy for different strengths of the spin-orbit
coupling. The edge behavior at $\alpha p_F = \omega_s$ predicted by Eq.(\ref{eq14}) is clearly
visible.
 When $\epsilon_F > \omega_s  \gg \alpha p_F$ the anisotropy
is nearly constant, which agrees both with our weak and strong field expansion.
\section{The conductivity within the Kubo formula approach}
In this section 	
we go beyond the Boltzmann equation approach in the relaxation-time
approximation and  calculate the conductivity via the 
Kubo formula. 
The starting point is the response kernel $Q_{ij}$ which  at zero temperature
is given by
\begin{eqnarray}
Q_{ij}(q)& = &{\rm i}e^2\sum_{{\bf p}}\int_{-\infty}^{\infty}
\frac{{\rm d}\epsilon}{2\pi}
{\rm Tr}\left[ j^i(p,q )G(p_+ )J^j(p,q)G(p_-)\right] \nonumber\\
&& -\frac{e^2N}{m}\delta_{ij}.
\end{eqnarray}
In the above we are using four vectors $p=(\epsilon , {\bf p} )$ and
$q=(\omega , {\bf q})$ with $p_{\pm}=p\pm q/2$;
the trace is over the spin.
In the presence of the spin-orbit coupling the bare current vertex differs
from the standard expression and is given by
\begin{equation}
\label{17}
{\bf j} =\frac{\bf p}{m}-\alpha{\mathbf \sigma}\wedge {\bf e}_z.
\end{equation}
The evaluation of the dressed vertex will be considered in section IV.
 The conductivity tensor is obtained by considering the limit
\begin{equation}
\label{18}
\sigma_{ij}=\lim_{\omega\rightarrow 0}
\frac{Q_{ij}(\omega , 0)}{{\rm i}\omega}.
\end{equation} 
By splitting the energy integration into  regions where the Green functions
have defined analytical properties, one may separate 
the kernel into two parts. In the first both the Green functions
are   retarded or advanced. This part cancels exactly  the diamagnetic term, i.e.,
\begin{equation}\label{eq19}
{\rm i}\sum_{{\bf p}}\int_{-\infty}^{\infty}\frac{{\rm d}\epsilon}{2\pi}
{\rm Tr}\left[ j^i(p)G(p)J^j(p)G(p) \right]-\frac{N}{m}\delta_{ij}=0.
\end{equation}
The cancellation is made explicit in appendix \ref{appendixb}
by exploiting the Ward  identity due to the gauge invariance.
The second term is given by
\begin{eqnarray}
\sigma_{ij}&=&\frac{e^2}{4\pi}\sum_{{\bf p}}
{\rm Tr}\left[ 2 j^i(p)G^R(p)J_{RA}^j(p)G^A(p) \right. \cr 
&-& j^i(p)G^R(p)J_{RR}^j(p)G^R(p)  \cr
&-& \left. 
j^i(p)G^A(p)J_{AA}^j(p)G^A(p)\right].
\label{eq20}
\end{eqnarray}
The dressed vertex depends whether it is connected to a pair of
retarded and advanced Green functions or 
a pair of Green functions with equal analytic properties.

\begin{figure}
\noindent
\begin{minipage}[t]{0.98\linewidth}
{\centerline{\epsfxsize=7cm\epsfbox{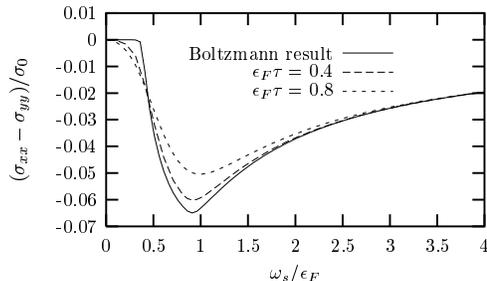}}}
\caption{$\Delta\sigma$ versus magnetic field 
for $\alpha p_F = 0.4$ and various scattering strengths. 
Scattering events are taken into account in the broadening of the energy levels as discussed in the
text.}
\label{figKubo}
\end{minipage}
\end{figure}

In order to make contact with the analysis of the previous section, we
temporarily make no assumption regarding a specific scattering mechanism
and simply take a (retarded) Green function
 of the form
\begin{equation}
\label{21}
( G^{R}_{\pm} )^{-1}({ \bf p}, \epsilon) 
 = \epsilon - E_{\pm }({\bf p}) + \mu +\I/ 2\tau
,\end{equation}
where the scattering time $\tau$ is of some unknown origin.
Let us also neglect the vertex corrections, i.e.
$J \to  j$. We are aware that neglecting vertex corrections often leads
to serious errors in the calculation of the conductivity,
but their  calculation  requires a 
specific scattering mechanism.
The conductivity is then given by
\begin{equation}
\label{22}
\sigma_{ij}=\frac{e^2}{\pi}\sum_{{\bf p}}
\sum_{a,b = \pm} 
\langle a |j^i(p)| b \rangle \Im ( G^R_b ({\bf p}) )
\langle b |j^j(p)| a \rangle \Im ( G^R_a ({\bf p}) )
.\end{equation}
In the limit of weak scattering $1/\tau \to 0$,
$\Im G^R$ becomes strongly peaked at the Fermi energy.
As a result intra-band contributions ($a=b$) dominate in this limit.
Neglecting then the inter-band terms ($a \ne b$) and with the approximation
\begin{equation}
\label{23}
\Im G^R_a  \Im G^R_a  \approx  \tau \pi \delta (\epsilon - E_a + \mu)
\end{equation}
one recovers  the expression for
the conductivity of the Boltzmann equation approach,
 when we identify $\tau_{\rm tr} $ with $\tau$.
In the general case with a finite scattering rate, the inter-band terms
and 
deviations of $\Im G^R \Im G^R$ from the delta function have to be taken
into account.
This is clearly demonstrated  in Fig.\ref{figKubo}.
Taking into account a finite $1/\tau$ the sharp edge of the anisotropy at 
$\omega_s = \alpha p_F$ disappears, although  the anisotropy curves do not change much even
in the strong scattering case where $\epsilon_F \tau = 0.8$. 

Going beyond this crude approximation in calculating the conductivity
we have to pay the price to make an
assumption on the scattering mechanism at play.
We assume elastic scattering, and we neglect quantum interference
corrections to the conductivity.
In the following we will first calculate the conductivity in the case of a pure s-wave scattering.
The results will be obtained in the framework of the self-consistent Born approximation.
Then we will allow an arbitrary $p$-dependence of the scattering potential. In that case
we will confine ourselves to the weak disorder limit.

\section{s-wave scattering}
In this section we consider a specific microscopic origin for 
the conductivity, namely elastic scattering.
For simplicity we assume a impurity potential with short range 
Gaussian correlations
\begin{equation}
\label{24}
\langle U({\bf x} ) U({\bf x }')\rangle  = {1\over 2\pi N_0 \tau}
\delta( {\bf x} - {\bf x}' )
.\end{equation}
We will calculate the conductivity in the framework of the self-consistent Born 
approximation, including the corresponding vertex corrections.

\begin{figure}
\noindent
\begin{minipage}[t]{0.98\linewidth}
{\centerline{\epsfxsize=7cm\epsfbox{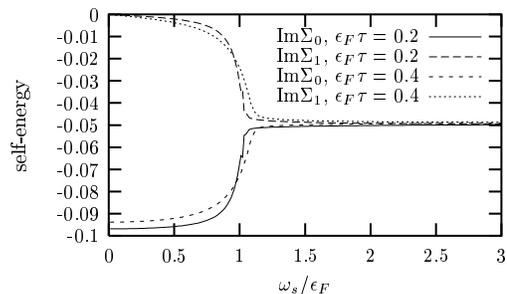}}}
\caption{
The imaginary part of the self-energy for $\alpha p_F = 0.4$ and $\epsilon_F \tau = 0.2$, $0.4$ as
a function of the magnetic field. The numbers for $\epsilon_F \tau = 0.4$ have been divided by two.
}
\label{figSelfenergy}
\end{minipage}
\end{figure}

The self-energy is then given by
\begin{equation}
\label{25}
\Sigma = {1\over 2\pi N_0 \tau} \sum_{\bf p} G(p)
.\end{equation}
To deal with the above 
 matrix equation we expand it in  Pauli
matrices. The Green function can be written as
$G=G_0\sigma_0 + G_1\sigma_1+G_2\sigma_2$ with
\begin{eqnarray}
\label{26}
G_0^{}   & =& 
\frac{1}{2}\left[G^{}_++G^{}_- \right]  \cr
G_1^{}   & =& -\frac{\omega_s -\alpha p_y - \Sigma_1^{} }{2\Omega^{}} 
\left[ G^{}_+-G^{}_-\right] \cr
G_2^{}   & =&-\frac{\alpha p_x }{2\Omega^{} }
\left[ G^{}_+-G^{}_-\right],
\end{eqnarray}
where $\Omega^{}({\bf p}) 
=\sqrt{(\alpha p_y -\omega_s+\Sigma_1^{})^2 +\alpha^2 p_x^2}$ and
\begin{equation}
\label{27}
G^{}_{\pm}=(\epsilon -p^2/2m +\mu \mp \Omega^{} - \Sigma_0^{} )^{-1}.
\end{equation}
Because the self-energy shares the matrix structure of the Green function,
it has no $\sigma_3$ component. Furthermore it turns out that $\Sigma_2 = 0$ 
always solves the self-consistency equation, due to the symmetry $p_x \to -p_x$.
As a result the self-energy has the form
$\Sigma_0 \sigma_0 +\Sigma_1 \sigma_1$.
To appreciate the meaning of the two self-energy components, 
we briefly consider the case with no spin-orbit coupling.
This is especially important for understanding the behavior at 
large magnetic field.
The real part of the self energy $\Sigma_0$ shifts the energy spectrum
by a constant. Since we have to adjust the chemical potential in order to keep
the particle number fixed, $\Re \Sigma_0$ can be safely  neglected.
The real part of $\Sigma_1$ may be re-absorbed into a renormalization of the Zeeman energy.
We then concentrate on the imaginary part, and find
\begin{equation}
\label{28}
\Im \Sigma^R = -\left(
 {1\over 2 \tau_0 } \sigma_0 + {1\over 2 \tau_1} \sigma_1 \right)
\end{equation}
with
\begin{eqnarray}
\label{29}
{1\over \tau_0 } & \approx  &{1\over \tau} {N_+ + N_- \over 2 N_0 } \\
{1\over \tau_1 } & \approx  &{1\over \tau} {N_- - N_+ \over 2 N_0 }.
\end{eqnarray}
The sum or difference 
$1/\tau_{\mp}= 1/\tau_0 \pm 1/\tau_1$ are nothing but the scattering times for the
two spin subbands.
For weak magnetic field $(\omega_s < \epsilon_F)$, the density of states in the
two subbands are identical and therefore
$ 1/\tau_\pm = 1/\tau_0 = 1/\tau$.
In the strong field limit $(\omega_s > \epsilon_F)$
the upper band is depopulated, so that
$ 1/\tau_1 = 1/\tau_0 = 1/2 \tau$.
The scattering rate in the empty band is zero due to the vanishing density of states.
Typical numerical results are shown in Fig.\ref{figSelfenergy}.
Although the general shape of the curves is as discussed above,  it is seen that for the disorder
strengths we consider there are already considerable modifications.

\subsection{The vertex corrections}
We now move to consider vertex corrections, following  the standard approach
in the literature (see for example \cite{agd}). Physically vertex corrections
are important for obtaining the correct momentum relaxation time,
which generally differs from the quasi-particle lifetime.
For the choice of a pure s-wave scattering potential, as in Eq.(\ref{24}), in the
absence of spin-orbit coupling, the momentum relaxation time coincides with
the quasi-particle lifetime.
This corresponds, diagrammatically, to the fact that 
there is no dressing of the current vertex with momentum independent impurity lines. 
In the presence of spin-orbit coupling this is no longer the case due to the presence
of extra terms in the current vertex of Eq.(\ref{17}).
The dressed current vertex is obtained by solving the equation
\begin{equation}\label{eq31}
J_{RA}^{x,y}= j^{x,y} + {1\over 2\pi N_0 \tau} \sum_{\bf p}  G^R J_{RA}^{x,y} G^A.
\end{equation}
A similar equation exists for the other energy ranges when one has
a pair of retarded  or advanced Green functions.
The strategy to solve the equation is to separate momentum dependent from 
momentum independent parts, expand the spin-dependent quantities in the Pauli matrix basis, 
and finally invert the resulting matrix equations.

We begin by separating the  $p$-dependent and the $p$-independent parts of the current operator:
\begin{eqnarray}j^{x,y} & = & p_{x,y}/m \mp \alpha \sigma_{2,1} \\
J^{x,y} &= &J_0^{x,y}(p)+ \Gamma^{x,y}.
\label{32}
\end{eqnarray}
Since we consider a $p$-independent impurity scattering the $J_0^{x,y} (p)$ part
is not dressed, i.e., 
\begin{equation}J_0^{x,y}(p)= p_{x,y}/m.\label{34}\end{equation}
The equation for the momentum independent part of the  current vertex reads
\begin{equation}
\Gamma^{x,y}_{s s'}=\gamma_{s s'}^{x,y}+
\frac{1}{2\pi N_0 \tau }\sum_{\bf p} \sum_{a b}G^R_{sa}\Gamma_{a b}^{x,y} G^A_{b s'}.
\label{35}\end{equation}
We included here the spin indices $s$, $s'$, $a$, and  $b= \uparrow, \downarrow$.
The quantities $\gamma^{x,y}$ are a sum of the bare vertices $\mp \alpha \sigma_{2,1}$ 
and a term which is generated by $p_{x,y}/m$,
\begin{equation}
\gamma^{x,y}_{ss'}= \mp \alpha (\sigma_{2,1})_{ss'} +\frac{1}{2\pi N_0\tau} \sum_{\bf p} \sum_a 
   {p_{x,y}\over m } 
   G^R_{s a} 
   G^A_{a s'}
.\label{36}\end{equation} 
In the Pauli matrix space  the vertex equation becomes:
\begin{equation}
\Gamma_{\rho}^i=\gamma_{\rho}^{i}+
\frac{1}{2}\sum_{\mu \nu \lambda} I_{\mu \nu }
{\rm Tr}
\left(\sigma_{\rho}\sigma_{\mu}\sigma_{\lambda}\sigma_{\nu}
\right)\Gamma_{\lambda}^i
\label{37}\end{equation}
with
\begin{equation}
I_{\mu \nu}= \frac{1}{2\pi N_0 \tau }
\sum_{\bf p} G^R_{\mu} G^A_{\nu}
.\label{38}\end{equation}
Some of the integrals $I_{\mu \nu}$
are zero and so the equations simplify. The final result is
\begin{eqnarray} \label{eq42}
\left( \begin{array}{c}\Gamma_0^y \\ \Gamma_1^y \end{array}\right)
&= &
\left( \begin{array}{c}\gamma^{y}_{0} \\ \gamma^{y}_{1} \end{array}\right) \cr
&+& 
\left( \begin{array}{ll}
I_{00}+I_{11}+I_{22} & I_{01}+I_{10}\\ 
I_{01}+I_{10}   & I_{00} +I_{11} -I_{22} \end{array}\right)
\left( \begin{array}{c}\Gamma_0^y \\ \Gamma_1^y \end{array}\right)
\end{eqnarray}
\begin{eqnarray} \label{eq43}
\left( \begin{array}{c}\Gamma_2^x \\ \Gamma_3^x \end{array}\right)
&= &
\left( \begin{array}{c}\gamma^{x}_{2} \\ 0  \end{array}\right) \cr
&+ &
\left( \begin{array}{ll}
I_{00}-I_{11}+I_{22} & \I(I_{01}- I_{10})\\ 
-\I( I_{01}- I_{10} )  & I_{00} -I_{11}+I_{22} \end{array}\right)
\left( \begin{array}{c}\Gamma_2^x \\ \Gamma_3^x \end{array}\right)
,\end{eqnarray}
where the bare vertices $\gamma^{x,y}$ are
\begin{eqnarray}
\gamma_0^{y} & =  & J^y_{00}+ J^y_{11}+J^y_{22} \cr
\gamma_1^{y} & =  & \alpha + J^y_{01}+J^y_{10} \cr
\gamma_2^{x} & =  & -\alpha + J^x_{02}+J^x_{20}
\end{eqnarray}
with
\begin{equation} \label{eq42a}
J_{\mu \nu}^i= \frac{1}{2\pi N_0 \tau }
\sum_{\bf p}  G^R_{\mu}(p_i/m) G^A_{\nu}.
\end{equation}

One might  try to determine the dressed 
vertices $\Gamma_\rho^{x,y}$ by solving the above matrix
equations. Whereas this is straightforward for $\Gamma_{2,3}^x$, the equation for
$\Gamma_{0,1}^y$ needs special care.
The analysis of Eq.(\ref{eq42}) shows that
 the matrix on the right-hand side of 
(\ref{eq42})
has an eigenvalue equal to one.
As a consequence  the matrix 
$$
\left( \begin{array}{ll}
1-(I_{00}+I_{11}+I_{22}) & -(I_{01}+I_{10})\\ 
-(I_{01}+I_{10})   &1-( I_{00} +I_{11} -I_{22}) \end{array}\right)
$$
cannot be inverted, 
and (\ref{eq42}) has no unique solution.
This property follows from  charge conservation and manifests in  the charge density vertex being singular
in the low frequency, long wavelength limit
(see also App.\ref{appendixb}).
To see explicitly this property 
 consider the imaginary part of the self-energy:
\begin{eqnarray}
\label{selfenergydifference}
\Sigma^R - \Sigma^A  &= & {1\over 2 \pi N_0 \tau} \sum_{\bf p}( G^R - G^A ) \cr
&= & {1\over 2 \pi N_0 \tau} \sum_{\bf p}G^R( \Sigma^R - \Sigma^A ) G^A .
\end{eqnarray}
By means of Eq.(\ref{selfenergydifference}) we can relate the imaginary part of the self-energy with
the integrals over pairs of retarded and advanced Green
functions $I_{\mu \nu}$.
Expanding in Pauli matrices we find
\begin{equation}
(\Sigma^R - \Sigma^A)_\rho =
{1\over 2 } \Tr (\sigma^\rho \sigma^\mu \sigma^\lambda \sigma^\nu )
I_{\mu \nu}( \Sigma^R - \Sigma^A )_\lambda
\end{equation}
which gives rise to the equation
\begin{equation}
\label{scatteringrates}
\left( \begin{array}{c}
1/\tau_0 \\
1/\tau_1 
\end{array} \right) =
\left( \begin{array}{cc}
I_{00} +I_{11} + I_{22} & I_{01}+I_{10} \\
I_{01}+I_{10}   & I_{00}+I_{11}-I_{22}
\end{array} \right)
\left( \begin{array}{c}
1/\tau_0 \\ 1/\tau_1  \end{array}
\right).
\end{equation}
This means that the vector of the scattering rates is an eigenvector of
the above matrix with eigenvalue $\lambda_0 = 1 $.
The second eigenvalue is then given by $\lambda_1 = 2 I_{00} + 2 I_{11}-1$.
We  now demonstrate that the 
eigenvector corresponding to $\lambda_1$ is proportional to $(\gamma_0^{y}, \gamma_1^{y})$.
We start from 
\begin{equation} \label{eq48}
{1\over 2 \pi N_0 \tau}\Tr \sum_{\bf p} {j^y}(G^R - G^A) = 0.
\end{equation}
We use again the relation
$G^R- G^A = G^R(\Sigma^R-\Sigma^A) G^A$
and arrive at
\begin{equation} \label{eq49}
{1\over 4\pi N_0 \tau}
\sum_{\bf p }{p^y \over m} \Tr\left[ \sigma^\mu
\left({1 \over \tau_0 }\sigma^0 +{1 \over \tau_1} \sigma^1 \right)
\sigma^\nu \right]
G^R_\mu G^A_\nu + {\alpha \over \tau_1} = 0
\end{equation}
with the final result that
\begin{eqnarray}\label{eq50}
{1\over \tau_0} 
(J_{00}^y+J_{11}^y+J_{22}^y )+{1\over \tau_1}(\alpha + J_{01}^y+J_{10}^y )
&= &0 
,\end{eqnarray}
so that the vector $(\gamma_0^y, \gamma_1^y)$ 
is perpendicular
to $(1/\tau_0, 1/\tau_1)$ and therefore necessarily is the second eigenvector of the matrix
on the right hand side of (\ref{scatteringrates}).
The solutions of (\ref{scatteringrates}) for the dressed vertex $(\Gamma_0^y, \Gamma_1^y)$ are then
given by $(\Gamma_0^y, \Gamma_1^y ) = (\gamma^y_0, \gamma^y_1)/(1-\lambda_1)$ plus
an arbitrary vector proportional to $(1/\tau_0, 1/\tau_1)$.
From (\ref{eq48}) and (\ref{eq49}) one 
observes that this arbitrary vector is irrelevant for the conductivity.
For simplicity we choose it to be zero.

Numerical results for the dressed vertices are shown in the figure.
Notice that Eqs.(\ref{scatteringrates}) and (\ref{eq50}) provide a nontrivial consistency check for the numerics.
\begin{figure}
\noindent
\begin{minipage}[t]{0.98\linewidth}
{\centerline{\epsfxsize=6.5cm\epsfbox{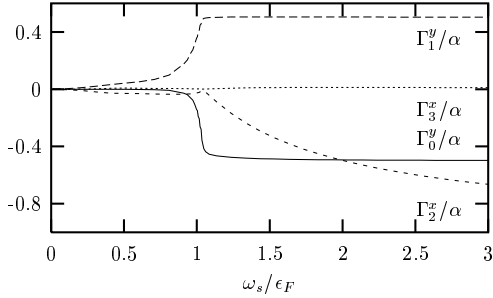}}}
\caption{ 
The dressed vertices $\Gamma_0 $ \dots $\Gamma_3$  in units of $\alpha$ as a function of the
magnetic field. Here we chose $\alpha p_F = 0.4 \epsilon_F$ and $1/(\epsilon_F \tau) = 0.2$.
Remember that the bare vertices are
$\gamma_0^y = \gamma_3^x = 0$ and $\gamma_1^y = -\gamma_2^x = \alpha$;
the asymptotic values in the strong field limit are
$\Gamma_1^y = - \Gamma_0^y =\alpha/2$, $\Gamma_2^x = -\alpha$, and $\Gamma_3^x = 0$. 
Notice that $\Gamma_2^x$ reaches the asymptotic value for high magnetic field only very slowly.
}
\label{figgamma}
\end{minipage}
\end{figure}
It is seen that the anomalous part of the current operator is strongly affected by
the vertex corrections. In particular for weak magnetic field the quantities $\Gamma_\mu$ 
become extremely small, so that the anomalous part of the current operator is canceled,
$J^i \approx p^i/m$.
In the limits
$\epsilon_F \tau \ll 1$ and $\omega_s=0 $ this cancellation  is exact as we now  show.
When $\omega_s =0$ the Green functions are
\begin{eqnarray}
G_0^{R(A)}   & =& 
\frac{1}{2}\left[G^{R(A)}_++G^{R(A)}_- \right]  \cr
G_1^{R(A)}   & =& \frac{ \sin (\theta )}{2} 
\left[ G^{R(A)}_+-G^{R(A)}_-\right] \cr
G_2^{R(A)}   & =&-\frac{ \cos (\theta )}{2}
\left[ G^{R(A)}_+-G^{R(A)}_-\right],
\end{eqnarray}
where 
\begin{equation}
G^{R(A)}_{\pm}=(\epsilon -p^2/2m \mp \alpha p \pm {\rm i}/2\tau)^{-1}
.\end{equation}
The first step is to evaluate the effective bare vertices entering the
expression for the conductivity, i.e., $\gamma^{i}$. This can be done
by evaluating the integrals we denoted by $J^i_{\mu\nu}$. Because the Green function
in the case $\omega_s =0$ has a simple angular dependence, one may
directly conclude that the integrals $J_{00}$, $J_{11}$, and $J_{22}$ are zero.
With the approximation $G^R_\pm G^A_\pm \approx 2 \pi \tau \delta(\epsilon - E_{\pm})$
the remaining integrals are readily found to be  $J_{01}+ J_{10}= -\alpha$ and $J_{02}+J_{20}=\alpha $
with the result that the vertices $\gamma^{i}$ and consequently the dressed vertices
$\Gamma_\mu$ are zero.

In the strong field limit $\omega_s \to \infty$ 
we find $J_x = p_x/m - \alpha \sigma_2 $, $J_y = p_y/m + \alpha (\sigma_1- \sigma_0 )/2$,
compare Fig.\ref{figgamma}.
Thus for $J_x$ the vertex corrections are absent but they seem to be present for $J_y$.
On the other hand as we noted before the solution of the vertex equation is not unique
and also $J_y = p_y/m + \alpha \sigma_1$ solves (\ref{eq31}), i.e. there are no vertex corrections to
the conductivity.
We will discuss the strong field limit in more detail in the next subsection.

\subsection{The conductivity}
\begin{figure}
\noindent
\begin{minipage}[t]{0.98\linewidth}
{\centerline{\epsfxsize=7cm\epsfbox{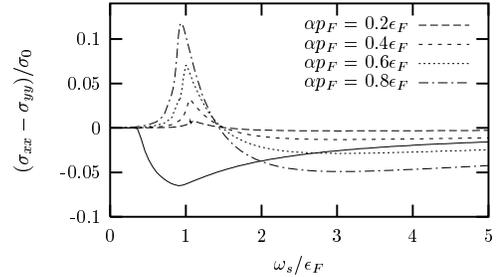}}}
\caption{Anisotropy in the magneto-conductance 
for $s$-wave impurity scattering, $1/(\epsilon_F \tau) = 0.2$ and various spin-orbit energies.
For comparison we included the results of the relaxation time approximation 
($\alpha p_F = 0.4 \epsilon_F$;
full line).  
For not too strong fields, the vertex corrections change the sign of the effect.}
\label{figAmrswave}
\end{minipage}
\end{figure}

\begin{figure}
\noindent
\begin{minipage}[t]{0.98\linewidth}
{\centerline{\epsfxsize=7cm\epsfbox{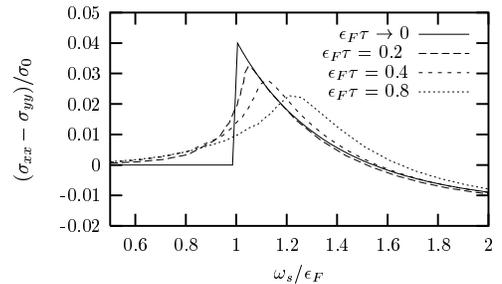}}}
\caption{Anisotropy in the magneto-conductance
for $\alpha p_F  = 0.4$ and varying the disorder. 
The curve in the limit of weak disorder was obtained within the approximation
described in the section on small angle scattering.}
\label{figAmrtau}
\end{minipage}
\end{figure}

Having discussed the vertex corrections we can now consider their effect on  the conductivity.
We do not find a strong magnetoresistance, compare also Fig.\ref{fig8} below.
Typical numerical results for the anisotropy in the conductivity are shown in Fig.\ref{figAmrswave}.
The most striking features are the change of sign at low magnetic fields
and the sharp structure around $\omega_s = \epsilon_F$.
In Fig.\ref{figAmrtau} it is seen that the structure around $\omega_s \approx \epsilon_F$ becomes more
pronounced the less disordered the system becomes. For $\epsilon_F \tau \ll 1$
a step in the anisotropy evolves at that energy.
The step is related to the van-Hove singularity in the density of states  at the band
edge of the upper band. Disorder smears out this singularity.
In order to have an understanding of these effects we will calculate the conductivity
analytically for strong magnetic fields.
The limit $\omega_s \to \infty $ is understood easily.
There is just one band occupied with the dispersion given by
$E_-({\bf p} ) = p^2/2m -\omega_s + \alpha p^y$.
The Fermi surface is a circle which is centered around $(0, -m\alpha)$. 
Shifting $p^y$ to $ p'^y= p^y + m\alpha$ 
the problem becomes rotational symmetric. The (bare) current operator in the lower band is
then simply of the standard form $j^{x,y} = p'^{x,y}/m$.
As a consequence the $p$-independent disorder potential does not renormalize the
current operator, and the conductivity is
$\sigma_{xx}= \sigma_{yy} = \sigma_0$. In particular there is no anisotropy in the conductance.

Lowering the magnetic field various contributions to the anisotropy appear.
Expanding the square roots in the energy dispersion,
the Green functions for the two subbands read
\begin{equation}
G_{\pm}^R(\omega , {\bf p})= \left(\omega +\mu \mp\omega_s 
-\frac{p_x^2}{2m_{\pm}}-\frac{p_y ^2}{2m} \pm \alpha p_y 
+{\rm i}\frac{1}{2\tau_{\pm}}\right)^{-1}
\end{equation}
where
\begin{eqnarray}
m_{\pm} &=&m\left( 1\pm\frac{m\alpha^2}{\omega_s}\right)^{-1}\\
\frac{1}{\tau_{\pm}}&=&\frac{1}{\tau_0}\mp\frac{1}{\tau_1}\left(1-\frac{1}{2}
\frac{\alpha^2p_x^2}{\omega_s^2} \right).
\end{eqnarray}
We discussed the effective mass anisotropy already in the framework of the
relaxation time approximation.
Allowing the scattering times in the two bands to be different,
the result of section \ref{secRelax}
is generalized to
\begin{equation}
\left( {\sigma_{xx} - \sigma_{yy} \over \sigma_0 }
\right)_{\rm mass}
= -{1\over 2}
{(\alpha p_F)^2 \over \omega_s \epsilon_F }
{\tau_- n_- - \tau_+ n_+ \over 
 \tau_- n_- + \tau_+ n_+ }.
\end{equation}
Further contributions to $\Delta \sigma$ are due to the anisotropic relaxation rate.
The scattering times $\tau_0$ and $\tau_1$ must be obtained by solving the 
self-consistent Born approximation. 
In the weak disorder limit
the van-Hove singularity at the  band edge of the upper band leads to a singular
 (=step) behavior
of the two scattering rates. Such singular behavior   is smeared out when disorder
 becomes stronger
(compare Fig.\ref{figSelfenergy}).
For simplicity we restrict the following discussion to the weak disorder case.
The angular dependent rate which we obtained by expanding the square root in the
energy dispersion
may also be obtained by considering the scattering probability
in the eigenstate basis. The latter
can be written as
\begin{equation} \label{eq56}
W_{{\bf p} {\bf p}'}^{\rm eff} =|U|^2\frac{1}{2}\left(
\begin{array}{c c}
1+{\bf e}_{\bf p}\cdot {\bf e}_{\bf p'} & 
1-{\bf e}_{\bf p}\cdot {\bf e}_{\bf p'}  \\
 1-{\bf e}_{\bf p}\cdot {\bf e}_{\bf p'} 
 &1+{\bf e}_{\bf p}\cdot {\bf e}_{\bf p'}
\end{array}
\right).
\end{equation}
The vector ${\bf e}_{\bf p}$ has been defined below Eq.(\ref{eq15}).
In the high field limit
\begin{equation}
W_{{\bf p} {\bf p}'}^{\rm eff} =|U|^2
\left(
\begin{array}{c c}
1 & 0  \\
0 & 1
\end{array}
\right)- |U|^2 {1\over 4}{ \alpha^2( p_x - p_x')^2 \over \omega_s^2} 
\left(
\begin{array}{r r}
1& -1 \\
-1 &1
\end{array}
\right),
\end{equation}
i.e. the effective scattering probability is anisotropic.
The anisotropy in the scattering rate is due to the term proportional to 
$(\alpha p_x)^2/\omega_s^2$ and is found as
$\delta (1/\tau_\pm) = \mp 2 \pi |U|^2 (\alpha p_x /2 \omega_s)^2 ( N_+ - N_- )$.
For $\omega_s < \epsilon_F$ the two density of states are practically identical,
so $1/\tau_\pm$ remains isotropic.
Only for $\omega_s > \epsilon_F$ the scattering rate is anisotropic and is a source for the
anisotropy in the conductivity, which we determine as
\begin{equation}
\left({\sigma_{xx} - \sigma_{yy} \over \sigma_0} \right)_{\tau_-} = {1\over 4} 
{(\alpha p_F)^2 \over  \omega_s^2 }.
\end{equation}
We now consider the vertex corrections.
To leading order in $1/\omega_s$ we need to calculate only the diagram 
with one impurity line. 
It is of the type
\begin{equation}\label{eq60}
\delta\sigma_{ii} =\frac{e^2}{2\pi}\sum_{{\bf p},{\bf p'}}
j^i({\bf p})G^R_{\bf p}G^A_{\bf p }W^{\rm eff}_{{\bf p}{\bf p'}}
G^R_{\bf p'}G^A_{\bf p'}j^i({\bf p}'),
\end{equation}
which one has to sum over the two bands.
Because the scattering
probability depends in the strong magnetic field limit 
only on the $x$ component of the momentum,
only $\sigma_{xx}$ is affected. We find 
by evaluating Eq.(\ref{eq60})
\begin{equation}
\left( {\sigma_{xx} - \sigma_{yy} \over \sigma_0 }\right)_{\rm vertex}
={1\over 2} { (\alpha p_F)^2\over \omega_s^2 }
\left({ n_+ - n_- \over n  }\right)^2.
\end{equation} 
When $\omega_s < \epsilon_F$
the vertex correction cancels with the contribution due to the
mass anisotropy since $(n_+-n_-)/n \approx \omega_s/\epsilon_F$, 
so that the 
anisotropy vanishes.
In the case $\omega_s > \epsilon_F$ the sum of the three terms we discussed gives
\begin{equation}
\Delta\sigma /\sigma_0 =
-{1\over 2} {(\alpha p_F)^2 \over \omega_s \epsilon_F } +
 {3\over 4}{(\alpha p_F)^2 \over \omega_s^2 }
\end{equation}
in full agreement with the numerical results of Fig.(\ref{figAmrswave}).
 
We conclude this section by noting  that the order of magnitude of the anisotropy 
in the microscopic calculation is the same as in the relaxation-time approximation.
Here however also the transport time becomes anisotropic.
The competition with the mass anisotropy kills $\Delta \sigma$ when $\omega_s < \epsilon_F$.
In the next section we will see that this cancellation 
is a peculiar consequence of the momentum independent impurity scattering.

\section{Small angle scattering} \label{secSmallAngle}
We will now drop the assumption of a 
$p$-independent impurity potential and consider the more general
case. 
For simplicity we will restrict ourselves to the weak disorder limit, where we
will neglect the disorder broadening of the energy levels, and terms which are related to the
off-diagonal matrix elements of the current operator (in the eigenstate basis).
The results obtained within these approximations are fully equivalent with
the Boltzmann equation approach, as it is discussed in Appendix B. 
In the energy state eigenbasis the dressing equation for the current vertex then reads
\begin{equation}
\label{61}
J_{ {\bf p} m}^{x,y} = j_{ {\bf p} m}^{x,y}
 + \sum_{ {\bf p}' m'} W^{\rm eff}_{ {\bf p} m; {\bf p}' m' }
 G^R_{{\bf p}'m'} G^A_{{\bf p}'m'} J_{{\bf p}' m'}^{x,y}
,\end{equation}
with
\begin{equation}
\label{62}
G^R_{{\bf p}m} G^A_{{\bf p}m}
\approx 2 \pi \tau_{{\bf p}m} \delta( \mu - E_m({\bf p} ) )
.\end{equation}
The lifetime is determined from the expression
\begin{equation}
\label{63}
1/\tau_{{\bf p} m} = 2 \pi \sum_{{\bf p}' m '}
 W^{\rm eff}_{{\bf p} m ; {\bf p}' m'} \delta(\mu - E_{m'}({\bf p}') )
.\end{equation}
We determine the dressed current operator with an expansion in 
eigenfunctions, following Ref.\cite{bhatt85}:
let $\phi^l_{ {\bf p} m} $ be the full set of eigenfunctions of the operator
\begin{equation}
\label{64}
\sum_{{\bf p}' m'} W^{\rm eff}_{{\bf p} m ; {\bf p}' m' }
G^R_{{\bf p}'m'} G^A_{{\bf p}'m'}  \phi^l_{{\bf p}' m'} = 
\lambda^l \phi^l_{ {\bf p} m},
\end{equation}
which are normalized according to
\begin{equation}
\label{65}
\sum_{{\bf p} m}
\phi^l_{{\bf p} m}G^R_{{\bf p}m} G^A_{{\bf p}m}  \phi^{l'}_{{\bf p} m} 
= \delta_{l l'}
.\end{equation}
The kernel entering the integral equation
(\ref{61}) may be expanded as
\begin{equation}
\label{66}
W^{\rm eff}_{{\bf p } m ; {\bf p}' m' } = \sum_l \lambda_l \phi^l_{{\bf p} m}
 \phi^l_{{\bf p}' m'}.
\end{equation}
One further notices that $1/\tau_{{\bf p} m}$ is an eigenfunction with eigenvalue one,
$\lambda^0 =1$ and $\phi^0_{{\bf p} m} \propto 1/\tau_{{\bf p} m}$.
Furthermore the bare current operator is perpendicular to this eigenfunction.
This can be shown in analogy with the discussion developed in the case of 
$s$-wave scattering in section IV.A.
With the expansion
\begin{equation}
\label{67}
j_{{\bf p} m}^{x,y} = \sum_l j_l^{x,y} \phi^l_{{\bf p} m};
\quad j_l^{x,y} = \sum_{{\bf p} m} 
j_{{\bf p} m}^{x,y}G^R_{{\bf p}m} G^A_{{\bf p}m}  \phi^{l}_{{\bf p} m}
.\end{equation}
the dressed current operator can be expressed as
\begin{equation}
\label{68}
J_{{\bf p} m }^{x,y} = \sum_{l \ne 0} \phi^l_{{\bf p} m} {1\over 1-\lambda_l } j_l^{x,y}
,\end{equation}
from which we finally obtain the conductivity by inserting it in Eq.(\ref{eq20}).
Before showing the results for the anisotropy and analyzing
the vertex corrections in the presence of small angle scattering, it is
instructive
to consider the zero magnetic field case first. As discussed in section IV 
for the case of
s-wave scattering, the vertex corrections 
have their most
dramatic effect on the anisotropy  in the limit of weak magnetic field. 
In the zero magnetic field limit, 
the structure of the dressed current vertex simplifies considerably.
The scattering probability $W_{{\bf p} {\bf p}'}$ and the effective scattering probability
become isotropic, i.e. they depend only on the angle between ${\bf p}$ and ${\bf p}'$.
One can then expand
\begin{equation}
W^{\rm eff} = W^0 + 2 W^1 \cos( \theta -\theta ' ) + \cdots
,\end{equation}
where $W^{\rm eff}$, $W^0$, and $W^1$ have still the two by two matrix structure. The
 lifetimes and
the dressed current operator are expressed as
\begin{eqnarray}
1/\tau_+ & =  & 2\pi N_+ W^0_{++} + 2\pi N_- W^0_{+-} \\
1/\tau_- & =  & 2\pi N_+ W^0_{-+} + 2\pi N_- W^0_{-+}
\end{eqnarray}
and
\begin{eqnarray}
\left( \begin{array}{c}
J_+^{x,y} \\ J_-^{x,y}
\end{array} \right) 
&= &
\left( \begin{array}{c}
j_+^{x,y} \\ j_-^{x,y}
\end{array} \right) \cr
&+&
\left( \begin{array}{cc}
2\pi N_+ \tau_+ W^1_{++} &  2\pi N_- \tau_- W^1_{+-} \\
2\pi N_+ \tau_+ W^1_{-+} &  2\pi N_- \tau_- W^1_{--}
\end{array} \right)
\left( \begin{array}{c} J_+^{x,y} \\ J_-^{x,y} \end{array} \right)
\end{eqnarray}
where $N_{\pm}(\epsilon )=N_0(1\mp m\alpha /\sqrt{(m\alpha)^2+2m\epsilon })$
and ${\bf j}_{\pm}= \nabla_{\bf p} E_{\pm}({\bf p})$ are the density of states and the bare current
vertices in the two bands.
For short range impurity potential, from Eq.(\ref{eq56}),  one finds
$W^0_{++} = \dots = 1/ 4 \pi N_0 \tau$, $\tau_\pm =\tau$, and
$W^1_{++} = - W^1_{+-} =\dots = 1/ 8 \pi N_0 \tau$.
Going through the algebra one may verify our earlier result that the vertex corrections 
cancel the anomalous velocity operator in this limit,
${\bf J}_\pm = {\bf p}_\pm/m$. Generally however this is not the case. 
In the extreme forward scattering limit in particular, one
can neglect the off-diagonal terms $W^1_{+-}$ and arrives at the standard expression for
single, isotropic bands
${\bf J}_{\pm} ={\bf j}_{\pm} \tau_{\rm tr}/\tau$
with
$1/\tau_{\rm tr} = N_\pm \int  \D \theta  W^{\rm eff}_{\pm \pm}(\theta) [ 1-\cos(\theta)] $.

Let us now consider a Gaussian scattering probability,
\begin{equation}
\label{69}
W_{ {\bf p } {\bf p}' } =  W  \exp( - r_0^2 ( {\bf p} - {\bf p}' )^2/2 ) 
.\end{equation}
The electron spin is assumed to be conserved, so this scattering potential has still to be
 transformed
to the energy eigenstate basis in order to obtain $W^{\rm eff}_{\bf p p'}$, see
 Eq.(\ref{eq56}).
The parameter $r_0$ controls the decay of the scattering potential. $r_0=0$ 
corresponds to pure s-wave
scattering, and with $r_0 p_F \gg 1$ we are in the strong forward scattering limit.
For simplicity the strength of the scattering potential 
and the screening parameter $r_0$
are here assumed to be independent of $\omega_s$.
Of course, in order to explain the large magnetoresistance such a dependence has to be
 taken into account. 
Suggestions for a microscopic origin of the magnetoresistance have been given, e.g., 
in Refs.\cite{dassarma00,dolgopolov00,gold00,herbut00}.
At the end of this section we will also consider the charged impurity scattering,
which was discussed in the absence of the spin-orbit coupling, e.g., in
 \cite{dolgopolov00,gold00}.

\begin{figure}
\noindent
\begin{minipage}[t]{0.98\linewidth}
{\centerline{\epsfxsize=7cm\epsfbox{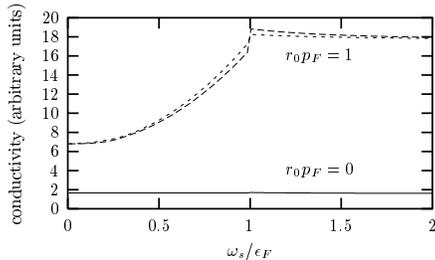}}}
\caption{Magneto-conductance
for $\alpha p_F  = 0.3$ and two different screening parameters $r_0$. 
Within the approximations in this section the conductivity is proportional to the inverse 
of the scattering strength, $W$, which we assume here to be independent of the magnetic
 field. 
For $s$-wave scattering ($r_0 = 0$) the anisotropy of the conductivity is not seen on the
 scale of the
plot. For $r_0 p_F=1$ we show the conductivities parallel and perpendicular to the magnetic
 field.
}\label{fig8}
\end{minipage}
\end{figure}

Although our main concern here is the anisotropy in the magnetoconductance,
it may be useful to begin by showing the numerical results
for the magnetoconductance itself (Fig.\ref{fig8}) for the model with a 
Gaussian scattering probability.
Given a scattering strength $W$, the conductivity increases with the parameter $r_0$ and
 with the magnetic field.
Both effects are understood easily.
The kinematics selects scattering processes with $q=|{\bf p}-{\bf p}'|\approx 2p_F$,
whereas the scattering potential of Eq.(\ref{69}) gives $q< 1/r_0$.
The condition for effective scattering becomes $r_0p_F \ll 1$.
 Increasing $r_0$ weakens the above condition  and the conductivity gets
enhanced.
As a function of the magnetic field the conductivity increases due to a similar reason.
The Fermi surface of the lower band grows and the effective condition $r_0 p_{F,-}\ll 1$
is no longer satisfied.

\begin{figure}
\noindent
\begin{minipage}[t]{0.98\linewidth}
{\centerline{\epsfxsize=7cm\epsfbox{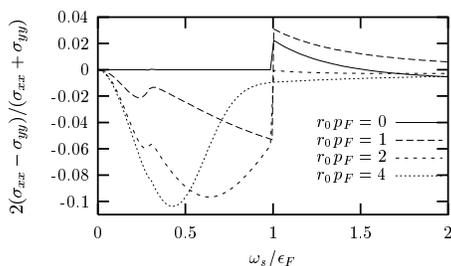}}}
\caption{Anisotropy in the magneto-conductance
for $\alpha p_F  = 0.3\epsilon_F$ and varying the screening parameter $r_0$.
The screening parameters $r_0 p_F=1,2,4$ correspond to 
$\tau_{\rm tr}/\tau \sim 2,8,30$ when they are determined for $\alpha p_F = \omega_s =0$.}
\label{figAmrShortRange}
\end{minipage}
\end{figure}

\begin{figure}
\noindent
\begin{minipage}[t]{0.98\linewidth}
{\centerline{\epsfxsize=7cm\epsfbox{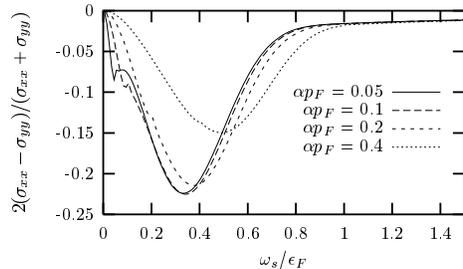}}}
\caption{Anisotropy in the magneto-conductance
for strong forward scattering ($r_0 p_F =4$) 
for different strengths of spin-orbit scattering. 
The curves for $\alpha p_F/\epsilon_F =0.2, 0.1, 0.05 $ are multiplied by $4,8,16$.
}\label{fig10}
\end{minipage}
\end{figure}

The numerically determined anisotropy in the magnetoconductance is shown in Fig.\ref{figAmrShortRange}.
Since now the conductivity itself is strongly magnetic field dependent we scale $ \Delta \sigma $
with $\sigma_{xx}+ \sigma_{yy}$ instead of $\sigma_0$.
One observes that the $p$-dependent scattering increases the overall 
size of the anisotropy up to a factor $\sim 5$ compared to the $s$-wave case. 
The sign of the anisotropy also changes. The peak anisotropy is still at magnetic fields of
 the
order of the Fermi energy, but there is a weak dependence on the scattering potential.
The anisotropy as a function of spin-orbit energy is shown in Fig.\ref{fig10}.
The overall amplitudes of the various curves are scaled in order to make them more comparable.
Also as a function of spin-orbit energy the position of the peak changes only weakly.

Finally let us consider charged impurity scattering.
This type of scattering may explain certain aspects of the 
magnetoresistance\cite{dolgopolov00}.
For a charged impurity situated in the plane the scattering potential is
\begin{equation}
U(q) = 2\pi e^2/[q \epsilon(q) ]
,\end{equation}
where $\epsilon(q)$ is the dielectric function.
We approximate it here as
\begin{equation}
{1\over \epsilon(q)} = {1\over 1+ (2\pi e^2/q)\chi(q) } 
\end{equation}
where $\chi(q)$ is the two-dimensional 
Lindhard function 
It has been argued in Ref.\cite{dolgopolov00} that the screening wave number 
$q_s = 4 \pi e^2  N(\epsilon_F)$ is for low electron density
large compared to the Fermi wavelength with the consequence that
\begin{equation}
U(q) \approx {1\over \chi(q)}
.\end{equation}
The scattering probability is approximated according to 
$W(q) \propto |U(q)|^2$.

\begin{figure}
\noindent
\begin{minipage}[t]{0.98\linewidth}
{\centerline{\epsfxsize=7cm\epsfbox{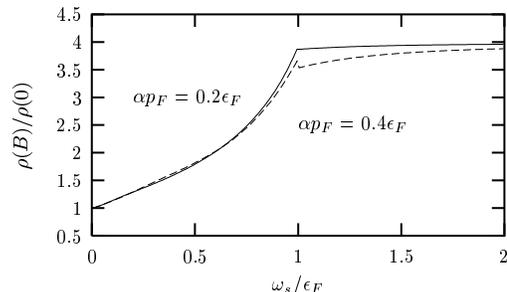}}}
\caption{
Resistivity versus in plane magnetic field for charged impurity scattering.
}\label{figGoldRho}
\end{minipage}
\end{figure}

In Fig.\ref{figGoldRho} we show the resistivity obtained with $\alpha p_F = 0.2$ and $0.4\epsilon_F$.
The resistivity increases by a factor of four from the zero field limit to very strong
magnetic
fields. In the absence of spin-orbit coupling the resistivity saturates when
 $\omega_s > \epsilon_F$, where
only one Fermi surface is left\cite{dolgopolov00}. 
In the presence of spin-orbit coupling there remains a weak increase of the resistivity
up to higher magnetic fields, since the depopulation of one band is no more equivalent with
the full spin polarization.

\begin{figure}
\noindent
\begin{minipage}[t]{0.98\linewidth}
{\centerline{\epsfxsize=7cm\epsfbox{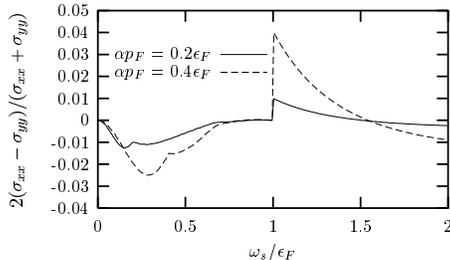}}}
\caption{
Anisotropy in the magnetoconductivity versus in plane magnetic field for charged
 impurity scattering.
The strength of the spin-orbit field is $\alpha p_F = 0.2, 0.4\epsilon_F$.
}\label{figGoldAmr}
\end{minipage}
\end{figure}

The anisotropy in the magnetoresistance for charged impurity scattering is shown
 in Fig.\ref{figGoldAmr}.
Within the approximation adopted 
the scattering probability  depends only weakly on the momentum.
Therefore results for the anisotropy are in the intermediate way between the $s$-wave 
and the strong
forward scattering limit. 

\section{Discussion}
We have calculated the Drude conductivity for a two-dimensional 
Fermi gas with spin-orbit coupling.
In the presence of an in-plane magnetic field the conductivity parallel
 and perpendicular to the
field are different.

Such an anisotropy has recently been observed in Si-MOSFETs\cite{pudalov00}. 
For the range of densities in the
experiment $n_s \approx 0.7 - 1.3 \times 10^{11}/{\rm cm}^2 $, 
the Fermi energy is $\epsilon_F \sim 5.1 - 9.5$K.
Taking $\alpha \approx 6\times 10^{-6}$Kcm which is reported in Ref.\cite{pudalov00}, 
the dimensionless spin-orbit
parameter  is of the order
$\alpha p_F/ \epsilon_F \sim 0.3 -0.7$. 
From our theory one expects a relative anisotropy in the
magnetoconductance of the order of several percent, 
in agreement to what is seen experimentally. 
The maximum anisotropy appears for magnetic fields of the
order $\omega_s \sim \epsilon_F$, so the peak energy scales with the density,
which again agrees with the experiments.
We found that the details of the effect depend on the scattering potential.
In the case of  pure $s$-wave scattering
the maximum anisotropy appears precisely at the magnetic field,
where the upper band depopulates.
The conductivity is larger when it is measured 
parallel to the magnetic field. On the other hand, at higher fields,
 $\omega_s > \epsilon_F$,
the conductivity is larger when measured perpendicular to the magnetic field. 
The sign of the anisotropy at low fields is at odds with the experiment. 
Allowing a finite range of the scattering potential, we have shown that the
 anisotropy sign changes.
In particular,
in the case of small-angle scattering both the sign and the peak position
 agree with what is
found experimentally. 
For the case of  charged-impurity scattering we
 found -- within the random-phase-approximation -- 
a rather complex pattern of the anisotropy.
There are two peaks with opposite sign, one peak near $\omega_s \sim \alpha p_F$ and
the other at $\omega_s \sim \epsilon_F$.
These results however should be taken with caution when comparing with experiments for
the low density electron gas.

Comparison with experiments in heterostructures is even more delicate due
the intrinsic crystallographic anisotropies. In both the experiments reported
in \cite{papadakis99,noh01} the anisotropy was measured along two crystallographic
directions with different mobilities and most likely with different operating scattering
mechanisms. At low densities  (close and across the MIT)
the  anisotropy shows $\Delta \sigma <0$, while at high densities there is
a change of sign as function of the magnetic field: $\Delta\sigma <(>)0$ for low (high)
fields. In our opinion, while a comparison with the sign of the anisotropy
contains the uncertainty related with the knowledge of the  scattering mechanism,
the fact that the maximum anisotropy at low and moderate fields scales with the density
appears as a robust feature in agreement with our theory.

In summary we think that the physics described in this paper may explain the anisotropic
magnetoresistance observed in Ref.\cite{pudalov00} in Si-MOSFETs, and 
perhaps that in heterostructures\cite{papadakis99,noh01}. At more general level,
our results give a further hint that the Drude-Boltzmann theory is a good starting point
 for the description
of the transport
properties of  the metallic phase of the 2D electron gas, even near the observed transition
 to an insulator.

We acknowledge fruitful discussions with C. Castellani and V. Falko.
This work was supported by the DFG through SFB 484 and by
EU Research Training Network program (Project Nr: RTN1 - 1999-00406).
 
\appendix
\section{Ward identity}
\label{appendixb}
In this appendix we derive for the sake of completeness the Ward identities for the
 system with spin-orbit coupling.
The generalized continuity equation is
\begin{equation}
\partial_t \rho({\bf x},t)+\partial_i {\rm I}^i({\bf x},t)=0
\end{equation}
where $i=x,y,z$. 
The density and the current operators are 
\begin{equation}
\rho({\bf x},t)=\psi^{\dagger}_{\alpha}({\bf x},t)
\psi_{\alpha}({\bf x},t)
,\end{equation}
\begin{eqnarray}
{\rm I}^i({\bf x},t)&=&\frac{{\rm i}}{2m}
\left[(\partial_i\psi^{\dagger}_{\alpha}({\bf x},t) )
\psi_{\alpha}({\bf x},t) \right. \cr
&-&\left. \psi^{\dagger}_{\alpha}({\bf x},t)
(\partial_i\psi_{\alpha}({\bf x},t))\right]\cr
 &-&\alpha\epsilon_{ijk}n_j \psi^{\dagger}_{\alpha}({\bf x},t)
 \sigma_{k,\alpha\beta}\psi_{\beta}({\bf x},t).
\end{eqnarray}
Define now the vertex functions
\begin{eqnarray}
   \Lambda^0_{ \alpha\beta}& =& 
   \langle T_t {\rho}^i({\bf x},t)\psi_{\alpha}({\bf x}',t'
)\psi^{\dagger}_{\beta}({\bf x}'',t'') \rangle \cr
\Lambda^i_{ \alpha\beta}   & =
& \langle T_t {\rm I}^i({\bf x},t)\psi_{\alpha}({\bf x}',t'
)\psi^{\dagger}_{\beta}({\bf x}'',t'') \rangle,  
\end{eqnarray}
for which the Ward identities read
\begin{eqnarray}
&&\partial_t \Lambda^0_{ \alpha\beta}+
\partial_i \Lambda^i_{ \alpha\beta}\cr
&=&\delta(t-t'')\delta ({\bf x}-{\bf x}''){\rm
i}G_{\alpha\beta}({\bf x}'t';{\bf x},t) \cr
&-&\delta(t-t')\delta ({\bf x}-{\bf x}')
\I  G_{\alpha\beta}({\bf x}t ;{\bf x}''t'').
\end{eqnarray}
After Fourier transform one gets
\begin{equation} \label{eqApp7}
\omega \Lambda^0_{ \alpha\beta}-q_i \Lambda^i_{ \alpha\beta}=
G_{\alpha\beta}(p_+, \epsilon_+)-G_{\alpha\beta}(p_-, \epsilon_-).
\end{equation}
In the static limit $\omega =0$ and letting ${\bf q}\rightarrow 0$ one gets
\begin{equation}
\Lambda^i_{\alpha \beta} 
=-G_{\alpha \sigma}(p,\epsilon )J^i_{\sigma \sigma '}(p,0) G_{\sigma ' \beta}(p,\epsilon)
= -\frac{\partial}{\partial  p^i} G_{\alpha \beta}(p, \epsilon).
\end{equation}
The above equation is sufficient to see the cancellation of the
diamagnetic term in the conductivity, see Eq.(\ref{eq19}).
\begin{eqnarray}
{\rm i}\sum_{{\bf p}}\int_{-\infty}^{\infty}\frac{{\rm d}\epsilon}{2\pi}
{\rm Tr}\left[ j^i(p)G(p)J^j(p)G(p) \right]-\frac{N}{m}\delta_{ij}\cr
={\rm i}\sum_{{\bf p}}\int_{-\infty}^{\infty}\frac{{\rm d}\epsilon}{2\pi}
{\rm Tr}\left[ j^i(p) 
\frac{\partial}{\partial  p^i} G(p)\right]-\frac{N}{m}\delta_{ij}\cr
=-{\rm i}\sum_{{\bf p}}\int_{-\infty}^{\infty}\frac{{\rm d}\epsilon}{2\pi}
{\rm Tr}\left[ \frac{\partial}{\partial  p^i}j^i(p) 
 G(p)\right]-\frac{N}{m}\delta_{ij}\cr
=-{\rm i}\frac{1}{m}\delta_{ij}\sum_{{\bf p}}\int_{-\infty}^{\infty}\frac{{\rm d}
\epsilon}{2\pi}
{\rm Tr}\left[  
 G(p)\right]-\frac{N}{m}\delta_{ij} 
=0 \end{eqnarray}
In the dynamic limit on the other hand $q=0$, $\omega \to 0$, Eq.(\ref{eqApp7})
becomes
\begin{equation}
\omega \Lambda^0_{\alpha \beta} = G^R_{\alpha \beta}
    - G^A_{\alpha \beta}
,
\label{eqApp9}\end{equation}
which can only be solved if the 
density vertex $\Lambda^0$ is singular in the zero frequency limit.
To see how this translates for the irreducible vertex $\Gamma$,
write $\Lambda^0_{\alpha\beta}=
G_{\alpha\sigma}\Gamma_{\sigma\sigma '}G_{\sigma '\beta}$ and perform
the ${\bf p}$-summation on both sides of Eq.(\ref{eqApp9}).
After decomposing in the Pauli matrices components, one gets
\begin{equation}
\omega\left(\left(
\begin{array}{c}
\Gamma_0 \\
\Gamma_1 
\end{array}
\right)-
\left(
\begin{array}{c}
\gamma_0\\
\gamma_1
\end{array}
\right)\right)={\rm i}
\left(
\begin{array}{c}
1/{\tau_0}\\
1/{\tau_1}
\end{array}
\right)
\label{eqApp10}
\end{equation}
where we have used Eqs.(\ref{eq42}).
From Eq.(\ref{eqApp10}) one sees that both $\Gamma_0$ and $\Gamma_1$
have to be singular in the zero frequency limit.

\section{Boltzmann equation}
We now briefly demonstrate the equivalence of the approach
in section \ref{secSmallAngle}
with the Boltzmann equation.
Since we have in mind to solve the Boltzmann equation in the presence
of a DC electric field, the distribution function $g_{{\bf p} \alpha}$ is chosen to depend
only on the wave vector ${\bf p}$. The Boltzmann equation including elastic scattering 
is then written
in the form
\begin{equation}
-{\rm e}{\bf E}\cdot {\bf \nabla}g_{{\bf p}\alpha}=
-\sum_{{\bf p}'\beta}Q_{{\bf p}\alpha, {\bf  p}'\beta}
\left( g_{{\bf p}\alpha}-g_{{\bf p}'\beta}\right).
\label{boltzmann}
\end{equation}
The Greek indices $\alpha$ and $\beta$ label the two spin subbands.
The scattering kernel, in the case of elastic scattering,
is related to the scattering probability by
\begin{equation}
 Q_{{\bf p}\alpha, {\bf p}'\beta}=2\pi 
 \delta\left( E_{{\bf p}\alpha}-E_{{\bf p}'\beta}\right)
 W^{\rm eff}_{{\bf p}\alpha, {\bf p}'\beta}
\end{equation}
A solution of the linearized Boltzmann equation is looked for in the form
\begin{equation}
g_{{\bf p}\alpha}=f(E_{{\bf p}\alpha})+\frac{\partial f}{\partial E}{\rm e}
{\bf E}\cdot {\bf u}_{{\bf p}\alpha}
\label{lb}
,\end{equation}
where the vector function ${\bf u}_{{\bf p}\alpha}$ obeys the integral equation
\begin{equation} \label{eqB4}
{\bf v}_{{\bf p}\alpha}=\sum_{{\bf  p}',\beta}
Q_{{\bf p} \alpha, {\bf p}'\beta}
\left( {\bf u}_{{\bf p}\alpha}-{\bf u}_{{\bf p}'\beta}\right).
\label{ie}
\end{equation}
Finally,
the electrical current density is given by
\begin{equation}
{\bf j}=- {e}\sum_{{\bf p},\alpha}{\bf v}_{{\bf p}\alpha}g_{{\bf p}\alpha}
\label{current}
,\end{equation}
which
becomes
\begin{equation}
{\bf j}={ e}^2 \sum_{{\bf p}\alpha}\left( -\frac{\partial f}{\partial
E}\right){\bf v}_{{\bf p}\alpha} ({\bf u}_{{\bf p}\alpha}\cdot {\bf E} )
\label{linearcurrent}
.\end{equation}
To make contact with the diagrammatic approach we write
\begin{equation}
{\bf u}_{{\bf p}\alpha}={\bf J}_{{\bf p}\alpha}\tau_{{\bf p}\alpha}.
\label{vertex}
\end{equation}
with
\begin{equation}
\tau^{-1}_{{\bf p}\alpha}=\sum_{{\bf  p}'\beta}
Q_{{\bf p}\alpha, {\bf  p}'\beta}.
\label{lifetime}
\end{equation}
The vector function ${\bf J}_{{\bf p}\alpha}$ is the renormalized
current vertex, which arises in the diagrammatic approach.
This is easily seen in the $T=0$ limit, by observing that
\begin{equation}
-\frac{\partial f}{\partial E}\rightarrow
\delta (\mu -E_{{\bf p}\alpha})\approx
\frac{1}{2\pi}\frac{1}{\tau_{{\bf p}\alpha}}G^R_{{\bf p}\alpha}
G^A_{{\bf p}\alpha}.
\label{approximation}
\end{equation}
Upon using 
${\bf u}_{{\bf p}\alpha}={\bf J}_{{\bf p}\alpha}\tau_{{\bf p}\alpha}$
and Eq.(\ref{approximation}), the integral equation (\ref{eqB4})
becomes
\begin{equation}
{\bf J}_{{\bf p}\alpha}={\bf v}_{{\bf p}\alpha}
+\sum_{{\bf  p}',\beta}
W^{\rm eff}_{{\bf p}\alpha, {\bf  p}'\beta}
G^R_{{\bf p}\alpha}
G^A_{{\bf p}\alpha}{\bf J}_{{\bf p}'\beta}
\label{vertexequation2}
\end{equation}
Hence the linearized Boltzmann equation is equivalent to the approach in section
\ref{secSmallAngle}, where we neglect lifetime broadening.

\end{multicols}
\end{document}